\documentclass[twocolumn,showpacs,prb]{revtex4}
\usepackage{graphicx}
\usepackage{dcolumn}
\usepackage{amsmath}
\usepackage{amssymb}

\makeatletter
\def\btt#1{\texttt{\@backslashchar#1}}
\DeclareRobustCommand\bblash{\btt{\@backslashchar}} \makeatother

\begin{document}

\title[Short Title]{Odd-frequency Pairs and Josephson Current through a Strong Ferromagnet}
\author{Yasuhiro Asano}
\affiliation{Department of Applied Physics, Hokkaido University, Sapporo 060-8628,
Japan}

\author{Yuki Sawa and Yukio Tanaka}
\affiliation{
CREST-JST and Department of Applied Physics, Nagoya University, Nagoya 464-8603, Japan}

\author{Alexander A. Golubov}
\affiliation{Faculty of Science and Technology, University of Twente, 7500 AE,
Enschede, The Netherlands }
\date{\today}

\begin{abstract}
We study Josephson current in superconductor / diffusive ferromagnet
/superconductor junctions by using the recursive Green function
method. When the exchange potential in a ferromagnet is sufficiently
large as compared to the pair potential in a superconductor, an
ensemble average of Josephson current is much smaller than its
mesoscopic fluctuations. The Josephson current vanishes when the
exchange potential is extremely large so that a ferromagnet is
half-metallic. Spin-flip scattering at junction interfaces
drastically changes the characteristic behavior of Josephson
current. In addition to spin-singlet Cooper pairs, equal-spin
triplet pairs penetrate into a half metal. Such equal-spin pairs
have an unusual symmetry property called odd-frequency symmetry and
carry the Josephson current through a half metal. The penetration of
odd-frequency pairs into a half metal enhances the low energy
quasiparticle density of states, which could be detected
experimentally by scanning tunneling spectroscopy. We will also show
that odd-frequency pairs in a half metal cause a nonmonotonic
temperature dependence of the critical Josephson current.
\end{abstract}

\maketitle

\section{introduction}
Ferromagnetism and spin-singlet superconductivity are competing
orders because the exchange potential breaks down spin-singlet
pairs. Spin-singlet pairs, however, do not always disappear under
the influence of an exchange potential. Long time ago
Fulde-Ferrell~\cite{fulde} and Larkin-Ovchinnikov~\cite{larkin}
discussed inhomogeneous spin-singlet superconductivity in the
presence of an exchange potential. It was shown that the
superconducting order parameter oscillates in real space because the
exchange potential shifts the center-of-mass momentum of a Cooper
pair. Similarly, a Cooper pair has been discussed in superconductor
/ ferromagnet (SF) and superconductor / ferromagnet / superconductor
(SFS)
junctions~\cite{buzdin,buzdin2,petrashov,ryazanov,kontos,golubov,bergeret,kadigrobov}.
These studies showed that a pairing function in a ferromagnet
changes its sign periodically in real space. As a consequence, SFS
junctions may undergo so-called 0-$\pi$ transition with varying
length of a ferromagnet or temperature.

Previous theoretical studies of the proximity effect in a
ferromagnet were mainly based on solving the quasiclassical Usadel
equations~\cite{usadel} valid when the exchange potential $V_{ex}$
is comparable to or smaller than the pair potential in a
superconductor at zero temperature $\Delta_0$. Cooper pairs can
penetrate into a ferromagnet within a short distance
$\xi_{h}=\sqrt{D/V_{ex}}$, where $D$ is the diffusion constant in a
ferromagnet. Thus, penetration of spin-singlet Cooper pairs into a
ferromagnet with large $V_{ex}$ would be impossible and the
Josephson coupling via such a strong ferromagnet would be
vanishingly small. A recent experiment~\cite{robinson}, however,
demonstrated the existence of Josephson coupling through a strong
ferromagnet with $V_{ex}\gg \Delta_0$. In addition to this, the
experiment~\cite{keizer} has even shown Josephson coupling in
superconductor /half metal / superconductor (S/HM/S) junctions. A
half metal is an extreme case of a completely spin polarized
material because its electronic structure is insulating for one spin
direction and metallic for the other. Thus one has to seek a new
state of Cooper pairs in a strong ferromagnet. The experiment by
Keizer et. al. has motivated a number of theoretical studies in this
direction~\cite{ya07l,braude,eschrig2,takahasi}.

Prior to the experiment~\cite{keizer}, Eschrig
\textit{et.al.}~\cite{eschrig} have addressed this challenging
issue. In the \emph{clean limit}, they have shown that $p$-wave
spin-triplet pairs induced by spin-flip scattering at a junction
interface can carry Josephson current. In practical S/HM/S
junctions, however, a half metal is close to the \emph{dirty limit}
in the diffusive transport regime; the elastic mean free path $\ell$
may be smaller or comparable to the coherence length and is much
smaller than the size of the half metal $L_N$. Thus, the effects of
the impurity potential on the Josephson current should be clarified
in a SFS junction consisting of a strong ferromagnet. In this paper,
we discuss the Josephson effect in SFS junctions for arbitrary
magnitude of $V_{ex}$. When $V_{ex}$ is much larger than $\Delta_0$,
an ensemble average of the Josephson current is much smaller than
its mesoscopic fluctuations~\cite{altshuler,zyuzin}. Fluctuations of
the pairing function in a ferromagnet is responsible for the large
fluctuations of Josephson current. The Josephson current vanishes in
S/HM/S in the absence of spin-flip scattering at junction
interfaces. Spin-flip scattering at junction interfaces drastically
changes the characteristic behavior of Josephson current and
properties of Cooper pairs in a ferromagnet. Spin-flip scattering
allows for the penetration of equal-spin-triplet Cooper pairs which
have unusual symmetry property called odd-frequency
symmetry~\cite{bergeret}. When the contribution of
equal-spin-triplet Cooper pairs to the Josephson current is
dominant, the self-averaging property of the Josephson current is
recovered. In particular in diffusive S/HM/S junctions, all Cooper
pairs in a half metal are in the odd-frequency equal-spin-triplet
pairing state~\cite{ya07l}. We also discuss local density of states
in a ferromagnet which reflects the existence of odd-frequency
Cooper pairs. A part of this study has been already published
elsewhere~\cite{ya07l}. Throughout this paper, we use the unit of
$\hbar=k_B=1$, where $k_B$ is the Boltzmann constant.

This paper is organized as follows. In Sec.~II, we explain the model
of SFS junctions on two-dimensional tight-binding lattice and the
method of calculation. The characteristic features of Josephson
current in SFS junctions are discussed in Sec.~III. In Sec.~IV, we
introduce spin-flip scattering at junction interfaces and discuss
symmetry properties of Cooper pairs in a ferromagnet. We propose an
experiment to observe odd-frequency pairs in SFS junctions based on
calculated results of local density of states in Sec.~V. The
conclusions are formulated in Sec.~VI.

\section{model}
\begin{figure}[tbh]
\begin{center}
\includegraphics[width=8.0cm]{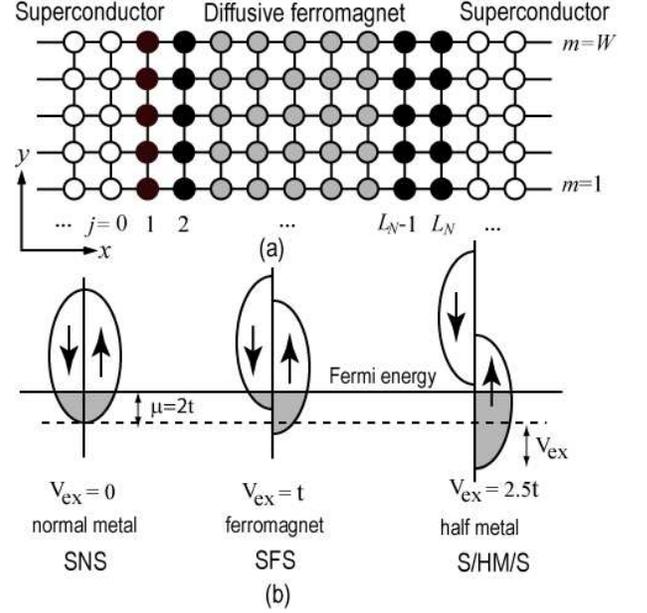}
\end{center}
\caption{ (Color online) (a) A schematic figure of a SFS junction on
tight-binding lattice. (b) Density of states for each spin direction.
The Josephson junction is of the SNS, SFS, and S/HM/S type for $V_{ex}/t=0$,
1 and 2.5, respectively.}
\label{fig1}
\end{figure}

Let us consider the two-dimensional tight-binding model as shown in
Fig.~\ref{fig1}(a). A vector $\boldsymbol{r}=j {\boldsymbol{x}}+
m{\boldsymbol{y}}$ indicates a lattice site, where
${\boldsymbol{x}}$ and ${\boldsymbol{y}}$ are unit vectors in the
$x$ and $y$ directions, respectively. A junction consists of five
segments: a ferromagnet ($3 \leq j \leq L_N-2$), two thin
ferromagnetic layers ( $ j=$ 1, 2, $L_N-1$, and $L_N$), and two
superconductors ($-\infty \leq j \leq 0$ and $L_N+1 \leq j \leq
\infty$). In the $y$ direction, the number of lattice sites is $W$
and we assume a periodic boundary condition. Electronic states in a
superconducting junction are described by the mean-field Hamiltonian
\begin{align}
H_{\textrm{BCS}}=& \frac{1}{2}\sum_{\boldsymbol{r},\boldsymbol{r}'}
 \left[ \tilde{c}_{\boldsymbol{r}}^\dagger\;  {h}_{\boldsymbol{r},\boldsymbol{r}'}
  \; \tilde{c}_{\boldsymbol{r}'}^{ }  -
\tilde{c}_{\boldsymbol{r}}^{t}\;  {h}^\ast_{\boldsymbol{r},\boldsymbol{r}'}
 \; \left\{ \tilde{c}_{\boldsymbol{r}'}^\dagger \right\}^{t} \right] \nonumber \\
 +&\frac{1}{2} \sum_{\boldsymbol{r},\boldsymbol{r}' \in \textrm{S}}
 \left[ \tilde{c}_{\boldsymbol{r}}^\dagger
\hat{\Delta}_{\boldsymbol{r},\boldsymbol{r}'}
\left\{\tilde{c}_{\boldsymbol{r}'}^\dagger\right\}^{t}
- \left\{\tilde{c}_{\boldsymbol{r}}\right\}^{t}
\hat{\Delta}^\ast_{\boldsymbol{r},\boldsymbol{r}'}
\tilde{c}_{\boldsymbol{r}'} \right], \label{bcs}\\
\hat{h}_{\boldsymbol{r},\boldsymbol{r}^{\prime }}=&
\left[ -t\delta _{|\boldsymbol{r}-\boldsymbol{r}^{\prime }|,1}
+(\epsilon _{\boldsymbol{r}}-\mu +4t)\delta _{\boldsymbol{r},\boldsymbol{r}^{\prime }}\right]
\hat{\sigma}_{0} \nonumber \\
& - \boldsymbol{V}(\boldsymbol{r})\cdot \hat{\boldsymbol{\sigma}}
\delta _{\boldsymbol{r},\boldsymbol{r}^{\prime }},\\
\hat{\Delta}_{\boldsymbol{r},\boldsymbol{r}'}=&
e^{i\varphi_j}
i \Delta \hat{\sigma}_2 \; \delta _{\boldsymbol{r},\boldsymbol{r}^{\prime }},  \\
\tilde{c}_{\boldsymbol{r}}=&\left( \begin{array}{c} c_{\boldsymbol{r},\uparrow} \\
c_{\boldsymbol{r},\downarrow}\end{array}\right),\;
\left\{ \tilde{c}_{\boldsymbol{r}} \right\}^t=(c_{\boldsymbol{r},\uparrow},
c_{\boldsymbol{r},\downarrow}),
\end{align}
where $c_{\boldsymbol{r},\sigma}^{\dagger}$
($c_{\boldsymbol{r},\sigma}^{ }$) is the creation (annihilation)
operator of an electron at $\boldsymbol{r}$ with spin $\sigma =$ (
$\uparrow$ or $\downarrow$ ), S in the summation means
superconductors, $\hat{\sigma}_j$ with $j=$1 - 3 are the Pauli
matrices, and $\hat{\sigma}_{0}$ is the $2\times 2$ unit matrix. The
hopping integral $t$ is considered among the nearest neighbor sites.
In a ferromagnet, on-site potential is given randomly in a range of
$-V_I/2 \leq \epsilon_{\boldsymbol{r}} \leq V_I/2$, where 
we take the probability distribution for $\epsilon_{\boldsymbol{r}}$ 
unifom on this interval, and $\epsilon_{\boldsymbol{r}}$ 
at different points  are uncorrelated.
The uniform exchange potential
in a ferromagnet is given by
$\boldsymbol{V}(\boldsymbol{r})=V_{ex}\boldsymbol{e}_{3}$, where
$\boldsymbol{e}_{l}$ for $l=1-3$ is a unit vector in spin space. The
Fermi energy $\mu$ is set to be $2t$ in a normal metal with
$V_{ex}=0$, while a ferromagnet and a half metal are respectively
described by $V_{ex}/t$ = 1 and 2.5 as shown in Fig.~\ref{fig1}(b).
Spin-flip scattering is introduced at $j=1,2$, $L_{N}-1$, and
$L_{N}$, where we choose
$\boldsymbol{V}(\boldsymbol{r})=V_{S}\boldsymbol{e}_{2}$. In a
superconductor, we take $\epsilon _{\boldsymbol{r}}=0$ and $\Delta$
is an amplitude of the pair potential in $s$-wave symmetry. The
macroscopic phases are given by $\varphi_j=\varphi_L$ in the left
superconductor and by $\varphi_j=\varphi_R$ in the right one.

The Hamiltonian is diagonalized by the Bogoliubov transformation,
\begin{align}
\left[
\begin{array}{c}
\tilde{c}_{\boldsymbol{r}} \\
\left\{\tilde{c}^\dagger_{\boldsymbol{r}}\right\}^{t}
\end{array}
\right]=& \sum_\lambda
\left[
\begin{array}{cc}
\hat{u}_\lambda(\boldsymbol{r}) & \hat{v}_\lambda^\ast(\boldsymbol{r}) \\
\hat{v}_\lambda(\boldsymbol{r}) & \hat{u}_\lambda^\ast(\boldsymbol{r})
\end{array}\right]
\left[ \begin{array}{c}
\tilde{\gamma}_{\lambda} \\
\left\{\tilde{\gamma}^\dagger_{\lambda}\right\}^{t}
\end{array}
\right], \label{bt}\\
\tilde{\gamma}_{\lambda}=&\left( \begin{array}{c} \gamma_{\lambda,\uparrow} \\
\gamma_{\lambda,\downarrow}\end{array}\right),
\end{align}
where ${\gamma}^\dagger_{\lambda,\sigma}$
(${\gamma}_{\lambda,\sigma}$) is the creation (annihilation) operator of a
Bogoliubov quasiparticle. The wave functions,
$\hat{u}_\lambda$ and $\hat{v}_\lambda$, satisfy the Bogoliubov-de Gennes
equation~\cite{degennes},
\begin{align}
\sum_{\boldsymbol{r}'}&
\left[ \begin{array}{cc}
\hat{h}_{\boldsymbol{r},\boldsymbol{r}'} &
\hat{\Delta}_{\boldsymbol{r},\boldsymbol{r}'}\\
-\hat{\Delta}^\ast_{\boldsymbol{r},\boldsymbol{r}'} & -\hat{h}^\ast_{
\boldsymbol{r},\boldsymbol{r}'} \end{array}
\right]
\left[ \begin{array}{c}
\hat{u}_\lambda(\boldsymbol{r}')  \\
\hat{v}_\lambda(\boldsymbol{r}')
\end{array}\right]
=  \left[ \begin{array}{c}
\hat{u}_\lambda(\boldsymbol{r})  \\
\hat{v}_\lambda(\boldsymbol{r})
\end{array}\right]\hat{E}_\lambda. \label{bdg}
\end{align}
The eigen value matrix $\hat{E}_\lambda$ is diagonal and depends on spin channels.
To solve the Bogoliubov-de Gennes equation, we apply the recursive Green function
 method~\cite{furusaki,ya01-1}.
In this method, we calculate the Matsubara Green function
\begin{align}
\check{G}_{\omega_n}(\boldsymbol{r},\boldsymbol{r}')
&= \sum_\lambda
\left[ \begin{array}{c}
\hat{u}_\lambda(\boldsymbol{r})  \\
\hat{v}_\lambda(\boldsymbol{r})
\end{array}\right]
[i \omega_n - \hat{E}_\lambda]^{-1}
\left[
\hat{u}_\lambda^\dagger(\boldsymbol{r}'),
\hat{v}_\lambda^\dagger(\boldsymbol{r}')
\right] \nonumber\\
+&\left[ \begin{array}{c}
\hat{v}^\ast_\lambda(\boldsymbol{r})  \\
\hat{u}^\ast_\lambda(\boldsymbol{r})
\end{array}\right]
[i \omega_n + \hat{E}_\lambda]^{-1}
\left[
\hat{v}_\lambda^t(\boldsymbol{r}'),
\hat{u}_\lambda^t(\boldsymbol{r}')
\right],\label{defg}\\
&=\left(
\begin{array}{cc}
\hat{g}_{\omega _{n}}(\boldsymbol{r},\boldsymbol{r}^{\prime }) & \hat{f}%
_{\omega _{n}}(\boldsymbol{r},\boldsymbol{r}^{\prime }) \\
-\hat{f}_{\omega _{n}}^{\ast }(\boldsymbol{r},\boldsymbol{r}^{\prime }) & -%
\hat{g}_{\omega _{n}}^{\ast }(\boldsymbol{r},\boldsymbol{r}^{\prime })%
\end{array}
\right), \label{deff}
\end{align}
where $\omega _{n}=(2n+1)\pi T$ is a Matsubara frequency, $n$ is an
integer number, and $T$ is the temperature. The Josephson current is
given by
\begin{equation}
J=-ietT\sum_{\omega _{n}}\sum_{m=1}^{W}\mathrm{Tr}\left[ \check{G}_{\omega
_{n}}(\boldsymbol{r}^{\prime },\boldsymbol{r})-\check{G}_{\omega _{n}}(%
\boldsymbol{r},\boldsymbol{r}^{\prime })\right]
\end{equation}
with $\boldsymbol{r}^{\prime }=\boldsymbol{r}+\boldsymbol{x}$. In this
paper, $2\times 2$ and $4\times 4$ matrices are indicated by $\hat{\cdots}$
and $\check{\cdots}$, respectively.

 In simulations, we first compute the Josephson current for a single sample with
a specific random impurity configuration. After calculating the
Josephson current over a number of different samples, an ensemble
average of the Josephson current and its fluctuations are obtained
as
\begin{align}
\langle J \rangle =& \frac{1}{N_s} \sum_{i=1}^{N_s} J_i, \label{jave}\\
\delta J =& \sqrt{ \langle J^2 \rangle - \langle J \rangle^2}, \label{jfluc}
\end{align}
where $J_i$ is the Josephson current in the $i$ th sample and $N_s$
is the number of samples. Strictly speaking, $N_s$ should be taken
to be infinity. In this paper, we increase $N_s$ until sufficient
convergence of $\langle J \rangle$ and $\delta J$ is obtained. In
the following, $N_s$ is typically taken to be 100-2000.

To study the characteristics of Cooper pairs in a ferromagnet, we
also analyze the anomalous Green function in Eq.~(\ref{deff}). The
pairing function is defined by the anomalous Green function and is
decomposed into four components,
\begin{equation}
\sum_{ \omega_c < \omega_n < \Delta_0}
\frac{1}{W}\sum_{m=1}^{W}\hat{f}_{\omega _{n}}(\boldsymbol{r},\boldsymbol{r}%
)=i\sum_{\nu =0}^{3}f_{\nu }(j)\hat{\sigma}_{\nu }\hat{\sigma}_{2},
\label{dec}
\end{equation}
where $\boldsymbol{r}=j\boldsymbol{x}+m\boldsymbol{y}$,
$\omega_c=0.01\Delta_0$ is a low energy cut-off and the pairing
functions are
averaged over whole lattice sites at $j$ before ensemble averaging.
In Eq.~(\ref{dec}),
$f_{0} (f_3)$ is the pairing function of spin-singlet (spin-triplet) pairs
with spin structure of $(\left\vert \uparrow \downarrow \right\rangle
-(+)\left\vert \downarrow \uparrow \right\rangle )/\sqrt{2}$.
The pairing functions of $\left\vert \upuparrows \right\rangle $
and $\left\vert\downdownarrows \right\rangle$
pairs are given by $f_{\uparrow \uparrow}=if_{2}-f_{1}$
and $f_{\downarrow \downarrow }=if_{2}+f_{1}$, respectively.

 The quasiclassical Green function method~\cite{eilenberger,usadel}
is a powerful tool to study the proximity effect when the pair potential
is much smaller than the Fermi energy.
The quasiclassical Green function, however, cannot be constructed in a
half metal because the Fermi energy for one spin
direction is no longer much larger than the pair potential.
On the other hand, there is no such difficulty in our method.
 These are advantages of the
recursive Green function method. Throughout this paper we fix the
following parameters: $L_{N}=74$, $W=25$, $\mu =2t$, and $V_{I}=2t$.
This parameter choice corresponds to the diffusive transport regime
in the N, F and HM layers~\cite{length}. The results presented below
are not sensitive to variations of these parameters.

\section{SFS junction without spin-active interface}
\begin{figure}[tbh]
\begin{center}
\includegraphics[width=8cm]{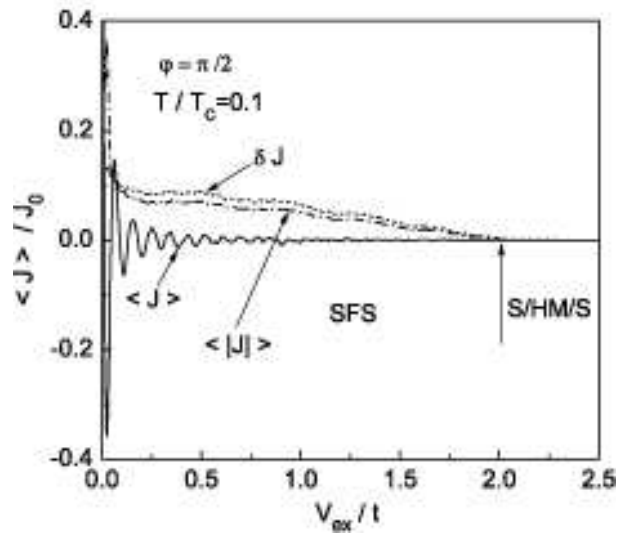}
\end{center}
\caption{ Josephson current is plotted as a function of the exchange
potential $V_{ex}$ for $V_{S}=0$ and $\Delta_0=0.01t$.
At $V_{ex}=2t$, a ferromagnet becomes half-metallic as
indicated by an arrow. The vertical axis is normalized by $J_0$ which is
an ensemble average of Josephson current at $V_{ex}=0$.
The number of samples used for averaging $N_s$ is 500.
}
\label{fig2}
\end{figure}

In this section, we do not consider spin-flip scattering at the
interfaces, (i.e., $V_S=0$). We first discuss the Josephson current
in SFS junctions as shown in Fig.~\ref{fig2}, where $T=0.1T_{c}$,
$\Delta_0=0.01t$, $T_{c}$ is the superconducting transition
temperature, and the phase difference across a junction
$\varphi=\varphi_L-\varphi_R $ is fixed at $\pi /2$.
The presented results are normalized by $J_{0}$ which is the
ensemble averaged of Josephson current in the superconductor /
normal metal / superconductor (SNS) junctions (i.e., $V_{ex}=0$).
The Josephson current oscillates as a function of $V_{ex}$ and
changes its sign almost periodically. The sign changes of $\langle
J\rangle $ correspond to the 0-$\pi $ transition of a SFS junction.
At the same time, the amplitude of $\langle J\rangle $ decreases
rapidly with increasing $V_{ex}$. For $V_{ex}>0.1t$, we should pay
attention to the relation $\langle J\rangle \ll \delta J$ which
means that the Josephson current is not a self-averaging quantity.
It is impossible to predict the Josephson current in a single sample
$J_{i}$ from $\langle J\rangle $ because $J_{i}$ strongly depends on
the microscopic impurity configuration. In fact, the Josephson
current flows in \emph{a single sample} even if $\langle J\rangle
=0$ at the transition points. Roughly speaking, $\langle J \rangle $
vanishes because half of samples are 0-junctions and the rest are
the $\pi$-junctions~\cite{zyuzin,ya01-2}. Since $\langle J\rangle
=0$, $\delta J$ approximately corresponds to the typical amplitude
of the Josephson current expected in a single sample. In
Fig.~\ref{fig2}, we also show $\langle |J| \rangle$, which agrees
well with $\delta J$ even quantitatively. The relation $\langle
J\rangle =0$ has different meaning for SFS and S/HM/S cases. In a
SFS junction, the fact that $\langle J\rangle =0$ at the transition
points is a result of ensemble averaging and the Josephson current
remains finite in a single sample. The characteristic temperature
and length of a ferromagnet at the $0-\pi$ transitions vary from one
sample to another. In S/HM/S junctions at $V_{ex}=2.5t$, however,
$\langle J\rangle =0$ means that the Josephson current vanishes even
in a single sample because $\delta J=0$ holds at the same time.
\begin{figure}[tbh]
\begin{center}
\includegraphics[width=8cm]{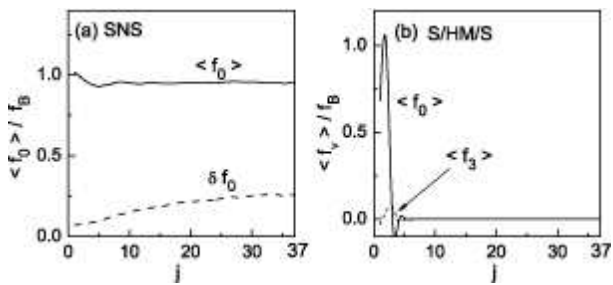}
\end{center}
\caption{ Pairing functions are plotted as a function of position
 $j$ for (a) SNS at $V_{ex}/t =0$ and (b) S/HM/S at $V_{ex}/t =2.5$, where
$V_{S}=0$, $\Delta_0=0.01t$ and $N_s=200$.
The vertical axis is normalized by a pairing function in a superconductor
$f_B$. }
\label{fig3}
\end{figure}
\begin{figure}[tbh]
\begin{center}
\includegraphics[width=8cm]{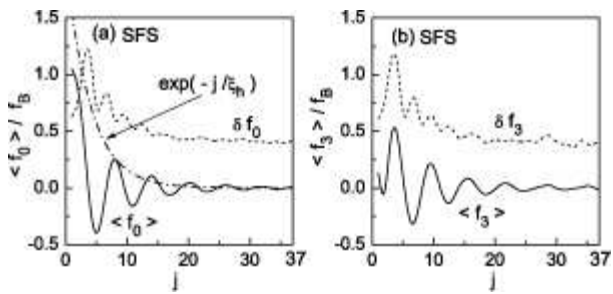}
\end{center}
\caption{ Pairing functions in SFS junctions
with $V_{ex}/t=1$, $V_{S}=0$, $\Delta_0=0.01t$ and $N_s=$500.
}
\label{fig4}
\end{figure}

The large fluctuations of Josephson current were discussed by Zyuzin
et. al.~\cite{zyuzin} by using the diagrammatic expansion. An
ensemble average of critical Josephson current and its fluctuations
have a relation for $W\gg L_N$
\begin{align}
\frac{\delta J}{ \langle J \rangle }\sim&
\sqrt{\frac{1}{W}}\frac{\exp(-L_N / \xi_T)}{\exp(-L_N / \xi_h)},\label{dj}\\
\sim & \sqrt{\frac{1}{W}} \exp\left\{ (\sqrt{V_{ex}} -\sqrt{\Delta_0})/\sqrt{E_{Th}} \right\},
\end{align}
where $E_{Th}=D/L_N^2$ is the Thouless energy, $\xi_{T}=\sqrt{D/2\pi
T}$, and $\xi_{h}=\sqrt{D/V_{ex}}$ is estimated to be about four
lattice constants (See also Appendix A). In the second line, we
replace $T$ by $\Delta_0$ because a measuring temperature must be
smaller than $T_c$. For a weak ferromagnet (i.e., $V_{ex} \lesssim
\Delta_0$), the ratio can be less than unity and the Josephson
current is a self-averaging quantity. On the other hand, in a strong
ferromagnet with $V_{ex} \gg \Delta_0$, the ratio becomes much
larger than unity. Thus the large fluctuation of Josephson current
is a robust feature of SFS junctions with $V_{ex} \gg \Delta_0$. The
only way to suppress fluctuations is taking the junction width
sufficiently large because fluctuations are a mesoscopic effect.

The origin of the large fluctuations in the Josephson current can be
understood by analyzing pairing functions of Cooper pairs. We plot a
pairing function of spin-singlet pairs $f_0$ in an SNS junction as a
function of position in a normal metal $j$ in Fig.~\ref{fig3}(a),
where $j=1$ and 37 correspond respectively to the junction interface
and the center of the normal metal. The pairing function is
calculated for $\varphi=0$ and is normalized by its bulk value in a
superconductor $f_B$. In SNS junctions, $\langle f_0 \rangle$ is a
real value and is almost constant as shown in Fig.~\ref{fig3}(a),
which means that spin-singlet Cooper pairs exist everywhere in the
normal metal. The pairing function for spin-singlet pairs in SFS
junctions is shown in Fig.~\ref{fig4}(a). An average $\langle
f_{0}\rangle $ decreases exponentially with $j$ according to $\exp
(-j/\xi_{h})$. At the same time, $\langle f_{0}\rangle $ oscillates
in real space and changes its sign. In addition to spin-singlet
pairs, opposite-spin-triplet pairs appear in a ferromagnet for
$V_{ex}\neq 0$. Since $f_3$ is a pure imaginary value, the imaginary
part of $\langle f_3 \rangle $ is plotted in Fig.~\ref{fig4}(b). The
behavior of $\langle f_{3} \rangle $ is qualitatively the same as
that of $\langle f_{0}\rangle$ in Fig.~\ref{fig4}(a). Thus
opposite-spin-triplet pairs also contribute to the Josephson
current. Both $\delta f_{0}$ and $\delta f_{3}$ remain finite at the
center of a ferromagnet $j=37$. Spin-singlet and
opposite-spin-triplet pairs penetrate deeply into a ferromagnet far
beyond $\xi_{h}$ even though $\langle f_{0}\rangle$ and $\langle
f_{3}\rangle$ are almost zero there. We numerically confirm the
relation $\delta f_{0}\propto e^{-j/\xi_{T}}$, in agreement with
Ref.~\onlinecite{zyuzin}.

\begin{figure}[tbh]
\begin{center}
\includegraphics[width=8cm]{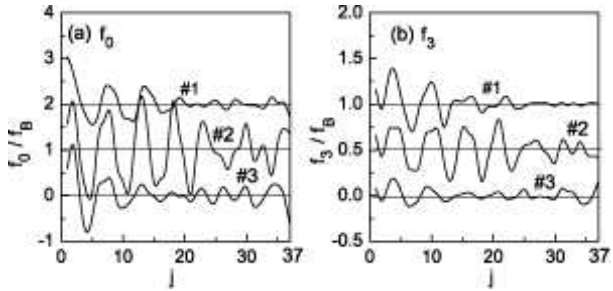}
\end{center}
\caption{ Pairing functions in three different samples of SFS junction at $V_{ex}/t=1$.
}
\label{fig5}
\end{figure}

In Fig.~\ref{fig5} (a) and (b), we show $f_0$ and $f_3$ in SFS
junctions for three samples with different impurity distribution.
The vertical axis is shifted as indicated by horizontal lines. The
pairing functions are in phase near the interface ($j\leq \xi_{h}$),
whereas they are out of phase far from the interface. Although the
pairing function in a sample has a finite value for $j > \xi_h$, an
ensemble average of them vanishes. Cooper pairs do exist in \emph{a
single sample} of ferromagnet even for $j\gg \xi_{h}$. Mesoscopic
fluctuations of the pairing function provide the origin of the large
fluctuations in the Josephson current. In S/HM/S junctions, as shown
in Fig.~\ref{fig3}(b), $\langle f_{0}\rangle $ and $\langle
f_{3}\rangle $ vanish for $j\gg 1$. We have also confirmed that
$\delta f_{0}=\delta f_{3}=0$ for $j\gg 1$ at the same time. Thus,
no Cooper pairs exist in a half metal for $V_S=0$.

\begin{figure}[tbh]
\begin{center}
\includegraphics[width=8cm]{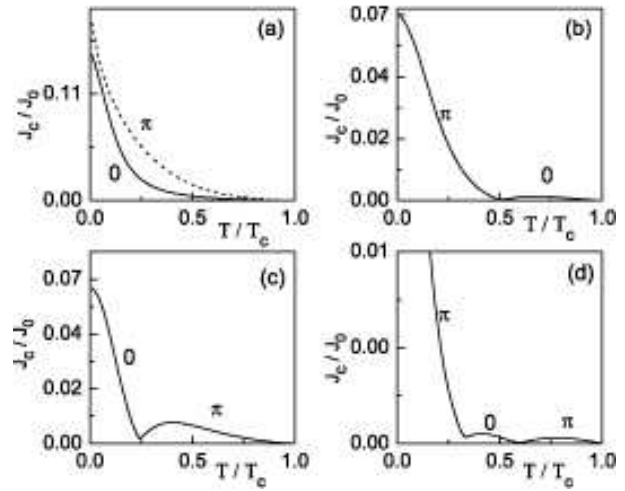}
\end{center}
\caption{ Critical current versus temperatures for five different samples of SFS junction,
where $V_{ex}/t=1$, $V_S=0$, and $\Delta_0/t=0.01$.
}
\label{fig6}
\end{figure}

 Since $\langle J \rangle \ll \delta J$, the temperature dependence of Josephson
current also depends on the impurity configuration. In
Fig.~\ref{fig6}, we show the Josephson critical current as a
function of temperature for five different samples, where the
critical current is estimated from the current-phase relation at
each temperature. The solid line in Fig.~\ref{fig6}(a) corresponds
to a SFS junction in the 0-state, where the critical current
monotonically increases with the decrease of temperature. On the
other hand, the broken line corresponds to a junction in the
$\pi$-state. In Fig.~\ref{fig6}(b), a junction undergoes the
transition from 0 to $\pi$ state when temperature decreases across
0.5$T_c$. On the contrary, the $0-$state is more stable than the
$\pi-$state at low temperatures in Fig.~\ref{fig6}(c). The Josephson
current is decomposed into a series of $J=\sum_{k=1}J_k
\sin(k\varphi)$. In Fig.~\ref{fig6}, $J_1=0$ characterizes the
0-$\pi$ transition temperature. At the transition temperature, the
critical current is not exactly zero because a higher harmonic such
as $J_2\sin(2\varphi)$ contributes to the Josephson current. Some
SFS junctions undergo the 0-$\pi$ transition twice as shown in
Fig.~\ref{fig6}(d). The temperature dependence of the critical
current in one sample can be very different from that in another
samples.

\section{SFS junction with spin-active interface}
The relation $\langle J\rangle \ll \delta J$ is a characteristic
feature of the Josephson current in diffusive SFS junctions with
$V_{ex} \gg \Delta_0$. This feature, however, is drastically changed
by spin-flip scattering at junction interfaces. In this section, we
study the Josephson current in the presence of spin-flip scattering,
(i.e., $V_S\neq 0$).
\begin{figure}[tbh]
\begin{center}
\includegraphics[width=8cm]{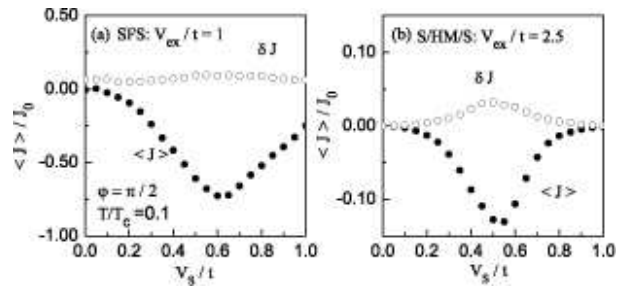}
\end{center}
\caption{ (a) Josephson current and its fluctuations for
$T=0.1T_{c}$, $\varphi =\pi /2$ and $N_s=200$ as a function of
interface spin-flip scattering $V_{S}$ for $V_{ex}/t=1$ and (b) for
$V_{ex}/t=2.5$.
The vertical axis is normalized by an ensemble average of
Josephson current at $V_{ex}=0$ and $V_S=0$.
}
\label{fig7}
\end{figure}
In Figs.~\ref{fig7} (a) and (b), we show $\langle J\rangle $ as a
function of the spin-flip potential $V_{S}$ for $V_{ex}/t=1$ and
2.5, respectively. In both cases (a) and (b), we find that $|\langle
J\rangle |\geq \delta J$ for $V_{S}\geq 0.3t$. The Josephson current
recovers the self-averaging property in the presence of spin-flip
scattering. Reasons can be found by analyzing the pairing functions
in a ferromagnet, as shown in Figs.~\ref{fig8} and \ref{fig9}, where
four pairing functions are plotted as a function of position $j$ at
$V_S/t=0.4$. In SFS junctions as shown in Fig.~\ref{fig8}(a),
equal-spin-triplet Cooper pairs penetrate into a ferromagnet by
spin-flip scattering at interfaces. Although averages of the pairing
function for opposite-spin pairs vanish at $j \sim 37$, their
fluctuations remain finite as shown in Figs.~\ref{fig8}(a) and (b).
Thus four types of Cooper pairs carry the Josephson current in a SFS
junction. In a S/HM/S junction, on the other hand, only
$\uparrow\uparrow$-pairs exist in a half metal as shown in
Fig.~\ref{fig9}(a) and (b). The pairing functions $\langle
f_{\downarrow \downarrow }\rangle $, $\langle f_0 \rangle $, and
$\langle f_3 \rangle $ vanish for $j \gg 1$. We note that
fluctuations of these pairing functions behave similar to their
averages. In both SFS and S/HM/S, $\langle f_{\uparrow \uparrow
}\rangle$ becomes much larger than $\delta f_{0}$ because the
exchange potential does not break down equal-spin-triplet Cooper
pairs and $f_{\uparrow \uparrow }$ does not suffer sign change in
real space. Thus the Josephson current becomes a self-averaging
quantity as shown in Figs.~\ref{fig7}(a) and (b).
\begin{figure}[tbh]
\begin{center}
\includegraphics[width=8cm]{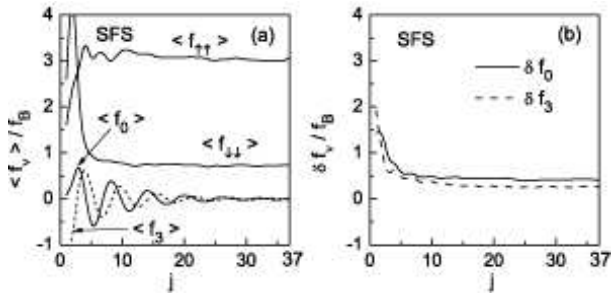}
\end{center}
\caption{ Pairing functions in SFS junctions are plotted as a junction of $j$.
Ensemble averages and some of their fluctuations are shown in (a) and (b), respectively.
The number of samples are taken to be 500.
 }
\label{fig8}
\end{figure}
\begin{figure}[tbh]
\begin{center}
\includegraphics[width=8cm]{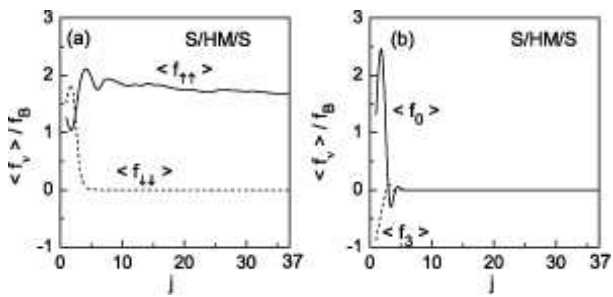}
\end{center}
\caption{ Pairing functions in S/HM/S junctions are plotted as a junction of $j$.
The number of samples are taken to be 200.
 }
\label{fig9}
\end{figure}

\begin{figure}[tbh]
\begin{center}
\includegraphics[width=8cm]{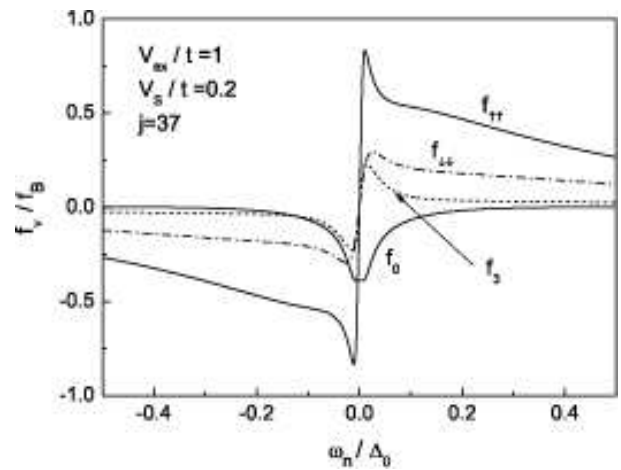}
\end{center}
\caption{ Pairing functions in a SFS junction are plotted as a junction of $\omega_n$
for $V_{ex}/t=1$ and $V_S/t=0.2$.
 }
\label{fig10}
\end{figure}

Here we address an unusual symmetry property of Cooper pairs in SFS
junctions. In Fig.~\ref{fig10}, we show four pairing functions in a SFS junction as a function of
$\omega_{n}$, where $j=37$, $V_{S}=0.2t$, $\varphi =0$, and $V_{ex}=t$.
Although the Green function at $\omega_n=0$ is not defined,
we put $f_{\uparrow \uparrow }=f_{\downarrow \downarrow }=f_3=0$ at $\omega_n=0$,
and connect results for positive $\omega_n$ with those for negative $\omega_n$.
 The pairing function
$ f_{0} $ is an even function of $\omega _{n}$, whereas $f_{\uparrow
\uparrow }$, $f_{\downarrow \downarrow }$, and $f_3$ are an odd
function of $\omega _{n}$~\cite{bergeret}. Since electrons obey
Fermi statistics, pairing functions must be antisymmetric under
interchanging two electrons,
\begin{equation}
\hat{f}_{\omega _{n}}(\boldsymbol{r},\boldsymbol{r}^{\prime })=-
\left[ {\hat{f}}_{-\omega _{n}}(\boldsymbol{r}^{\prime },\boldsymbol{r})\right]^t,
\label{pauli}
\end{equation}
where $[\hat{f}]^t$ denotes the transpose of $\hat{f}$ meaning the
interchange of spins. It is well known that ordinary even-frequency
pairs are classified into two symmetry classes: spin-singlet
even-parity symmetry and spin-triplet odd-parity one. In the former
case, the negative sign on the right hand side of Eq.~(\ref{pauli})
arises due to the interchange of spins, while in the latter case due
to $\boldsymbol{r}\leftrightarrow \boldsymbol{r}^{\prime }$. In the
present calculation, all components on the right hand side of
Eq.~(\ref{dec}) have $s$-wave symmetry. This is because the pairing
functions are isotropic in both real and momentum spaces due to
diffusive impurity scatterings~\cite{tanaka06}. As a result,
$f_{\uparrow \uparrow }$, $f_{\downarrow \downarrow }$, and $f_3$
must be an odd function of $\omega _{n}$ to obey Eq.~(\ref{pauli}).
The fraction of odd-frequency pairs depends on parameters such as
the exchange potential and the spin-flip potential. As shown
Fig.~\ref{fig3}(a), all Cooper pairs have even-frequency symmetry in
SNS junctions at $V_{ex}=0$ and $V_S=0$. Even- and odd-frequency
pairs have almost same fraction in SFS junctions at $V_{ex}=t$ and
$V_S=0$ as shown in Fig.~\ref{fig4}. In the presence of spin-flip
potential, odd-frequency pairs become dominant as shown in
Fig.~\ref{fig8}. In particular, all Cooper pairs have odd-frequency
symmetry in S/HM/S junctions as shown in Fig.~\ref{fig9}. The
Josephson current in Fig.~\ref{fig7}(b) is carried purely by
odd-frequency pairs in S/HM/S junctions.

The results in Fig.~\ref{fig7} show that the amplitude of Josephson
current first increases with the increase of $V_S$ then decreases.
Here we discuss the analytical expression of the Josephson current
in S/HM/S junction at $T=0$,
\begin{align}
&\langle J \rangle = - J_1 \nonumber \\
&\times
\left[ \left(\boldsymbol{V}_L\cdot\boldsymbol{V}_R - {V}_L^{(3)} {V}_R^{(3)}\right)\sin \varphi
 + \boldsymbol{e}_3 \cdot (\boldsymbol{V}_L\times\boldsymbol{V}_R) \cos\varphi\right],\label{jhm}\\
&J_1= \frac{7 \zeta(3)}{\pi} eE_{Th} \; g_N b^2>0,\\
&b=\frac{1}{2}\int_0^{\pi/2}\!\!\!\!d\gamma\;
\frac{\cos^5\gamma}{\left(V_S^2+\frac{1}{4}\right)\cos^4\gamma - V_S^2 \cos^2\gamma
+V_S^4}.
\end{align}
Here $\boldsymbol{V}_R=\sum_{k=1}^{3} V_R^{(k)}\boldsymbol{e}_k$ and
$\boldsymbol{V}_L=\sum_{k=1}^{3} V_L^{(k)}\boldsymbol{e}_k$ are the
dimensionless magnetic moments at the right and left junction
interface, respectively. We assume that
$|\boldsymbol{V}_R|=|\boldsymbol{V}_L|=V_S$ and $(e^2/h)g_N$ is the
normal conductance of a half metal. Details of derivation are
discussed in Appendices A and B. To compare Eq.(\ref{jhm}) with the
results in Fig.~\ref{fig7}(b), we choose
$\boldsymbol{V}_R=\boldsymbol{V}_L=V_S\boldsymbol{e}_2$. We also
note that the magnetic moment in a half metal is
$\boldsymbol{V}_{ex}=V_{ex}\boldsymbol{e}_3$. For $V_S \ll 1$, the
amplitude of the Josephson current increases with $V_S^2$ because
$b=2$ and $\boldsymbol{V}_L\cdot\boldsymbol{V}_R - {V}_L^{(3)}
{V}_R^{(3)} = V_S^2$. In this case, spin-flip scattering assists the
Josephson current. For large $V_S$, on the other hand, the Josephson
current decreases proportionally to $V_S^{-6}$ because the spin-flip
potential acts like a potential barrier and suppresses the
transmission probability of the interface. The Josephson current
shows reentrant behavior as shown in Fig.~\ref{fig7}. The Josephson
current in Fig.~\ref{fig7}(b) is calculated at $\varphi=\pi/2$. The
results indicate that the S/HM/S junction is a $\pi$-junction. This
conclusion depends on the direction of the magnetic moments at the
spin-flip interfaces. In the case of $\boldsymbol{V}_L =
\boldsymbol{V}_R$, the Josephson current in Eq.~(\ref{jhm}) is
proportional to $-J_1 \sin\varphi$ in agreement with
Fig.~\ref{fig7}(b). In the case of antiferromagnetic alignment,
$\boldsymbol{V}_L = -\boldsymbol{V}_R$, the junction is in the
0-state. Thus we conclude that the stability of the $0$-state and
that of the $\pi$-state depend on the alignment of the magnetic
moments at the two interfaces~\cite{volkov,houzet1}. This feature
indicates a new direction to controlling the $\pi$-phase shift by
using ferromagnetic materials. For $ \boldsymbol{V}_{ex} \cdot
(\boldsymbol{V}_L \times \boldsymbol{V}_R) \neq 0$, the Josephson
current flows even at $\varphi=0$ because such spin configuration
breaks the chiral symmetry of a junction. We have numerically
confirmed the Eq.~(\ref{jhm}).

\begin{figure}[tbh]
\begin{center}
\includegraphics[width=8cm]{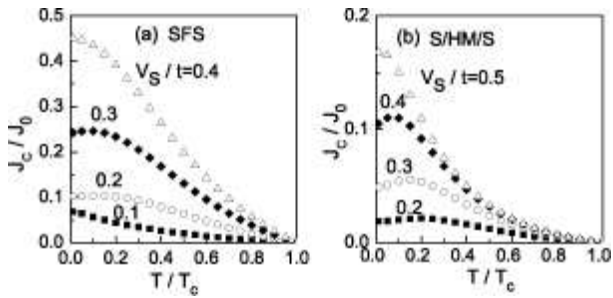}
\end{center}
\caption{ Temperature dependence of critical Josephson current in SFS (a) and
S/HM/S (b) junctions. $N_s$ is taken to be 200.
 }
\label{fig11}
\end{figure}

 In the end of this section, we discuss the temperature dependence of
the Josephson critical current in SFS and S/HM/S junctions in Fig.~\ref{fig11},
 where we choose $\Delta_0=0.005t$ in connection with the density of states
in the next section.
The Josephson current has almost a sinusoidal current-phase relationship.
The critical current for $V_S/t=0.2$ and 0.3 in a SFS junction first increases
with the decrease of temperature then decreases as shown in Fig.~\ref{fig11}(a).
Such reentrant behavior is seen more clearly in a S/HM/S junction as shown for
 $V_S/t=0.2$, 0.3 and 0.4 in Fig.~\ref{fig11}(b).
In a Josephson junction consisting of conventional $s$-wave
spin-singlet superconductors, such reentrant behavior is very
unusual. This behavior has also been reported in
Ref.~\onlinecite{eschrig2}. The results for $V_S/t=0.4$ in SFS and
$V_S/t=0.5$ in S/HM/S junctions, on the other hand, show a monotonic
temperature dependence.

\section{density of states}
\begin{figure}[tbh]
\begin{center}
\includegraphics[width=8cm]{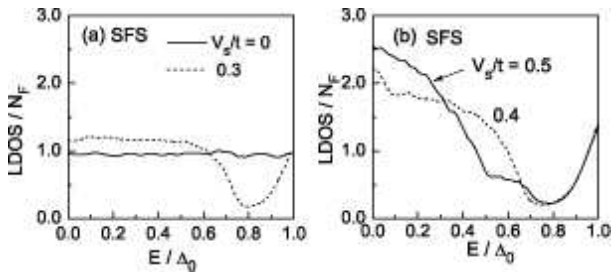}
\end{center}
\caption{ Local density of states at $j=37$ in a ferromagnet of SFS junction,
where $N_s=2000$.
 }
\label{fig12}
\end{figure}

\begin{figure}[tbh]
\begin{center}
\includegraphics[width=8cm]{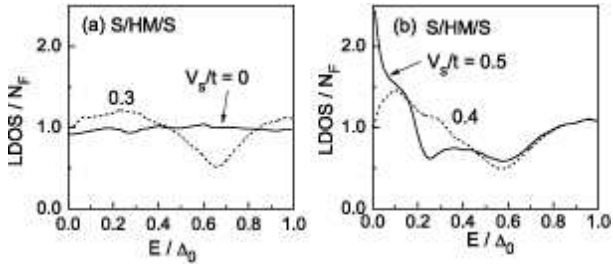}
\end{center}
\caption{ Local density of states at $j=37$ in a half metal of S/HM/S junction,
where $N_s=2000$.
 }
\label{fig13}
\end{figure}

The proximity effect changes the low energy spectra of a
quasiparticle in a normal metal. In a SNS junction, it is well known
that the penetration of usual even-frequency spin-singlet $s$-wave
Cooper pairs suppresses the quasiparticle density of states for
$E<E_{Th}$. This suppressed density of states is called minigap. In
this section, we discuss the proximity effect of odd-frequency pairs
on the quasiparticle density of states. In our method, the density
of states is given by
\begin{align}
N(E,j)=-\frac{1}{\pi }\frac{1}{W}
\sum_{m=1}^{W}\text{Im}\text{Tr}\check{G}_{E+i\gamma }(\boldsymbol{r},
\boldsymbol{r}),
\end{align}
where $\gamma $ is a small imaginary part.
In Fig.~\ref{fig12}, we show the local density of
states (LDOS) at $j=37$ in SFS junctions,
where $\varphi=0$, and $\gamma=0.1\Delta_0$.
The results for S/HM/S junctions are presented in Fig.~\ref{fig13}.
The LDOS is normalized by its value at $E=1.2\Delta_0$.
Here we choose $\Delta_0=0.005t$ so that $E_{Th}\sim 0.3\Delta_{0}$ is slightly
smaller than $\Delta_0$.
In the absence of spin-flip scattering, the ensemble average of LDOS is almost constant
in both Figs.~\ref{fig12} and ~\ref{fig13}.
At $V_S/t=0.3$, the penetration of odd-frequency pairs enhances LDOS for $E<0.5\Delta_0$.
On the other hand, LDOS is suppressed around $E\sim 0.8\Delta_0$ because of a sum rule
for the density of states (i.e., $\int dE\; N(E,j)=const.$).
The low energy spectra of LDOS increase with increasing $V_S/t$ as shown in
Figs.~\ref{fig12}(b) and \ref{fig13}(b).
At $V_S/t=0.5$, LDOS has a peak at $E=0$.
Thus the penetration of odd-frequency pairs enhances the quasiparticle density of states
for $E<E_{Th}$.
This tendency is just opposite to the minigap structure due to penetration of
even-frequency pairs.
The shape of the zero-energy peak in Figs.~\ref{fig12}(b) and \ref{fig13}(b) is almost
independent of the position in a half metal.
The peak is much stronger than the enhancement of the
LDOS found in weak ferromagnets~\cite{buzdin,golubov,fogelstrom,kontos2,yokoyama}.
In such SF junctions~\cite{kontos2}, the LDOS has an oscillatory peak/dip
structure at $E=0$, which rapidly decays with the distance from the SF interface.
Therefore, the large peak at $E=0$ in the LDOS is a robust and direct evidence of
the odd-frequency pairing in a ferromagnet.
Scanning tunneling spectroscopy (STS) could be used to detect such a peculiar pairing state.

 As shown in Fig.~\ref{fig13}, however, the penetration of odd-frequency pairs
does not always give rise to a zero-energy peak in LDOS.
The results for $V_S/t=0.3$ and 0.4 have a broad peak at a finite energy smaller
than $E_{Th}$.
This situation is slightly different from the large zero-energy peak in a normal metal
due to the penetration of odd-frequency pairs from spin-triplet odd-parity
superconductors~\cite{yt04,yt05r,ya06l,ya07-2}.
In a spin-triplet superconductor junction,
 LDOS in a normal metal always has a large zero-energy peak
because a midgap Andreev resonant state~\cite{yt95l} assists the zero-bias peak.
In ferromagnetic junctions, on the other hand, such a quasiparticle state is
absent at the junction interface.
The broad peak structure in the LDOS is responsible for
the nonmonotonic temperature dependence
of the critical current in Fig.~\ref{fig11}(b).
At high temperatures, quasiparticle states around the peak contribute to the Josephson
current. At low temperatures, however, such quasiparticle states cannot contribute to
the Josephson current~\cite{ya02-4}.
We conclude that odd-frequency pairs could also be confirmed by measuring the
dependence of the critical current on temperature.

\section{conclusion}
In conclusion, we have studied the Josephson effect in superconductor /
diffusive ferromagnet / superconductor (SFS) junctions by using the
recursive Green function method.
When the exchange potential in a ferromagnet is much larger than the
pair potential in a superconductor,
the Josephson current is not a self-averaging quantity.
This is because spin-singlet Cooper pairs
penetrating into a ferromagnet far beyond $\xi_{h}$ cause large
fluctuations of the pairing function.
As a consequence, the temperature dependence of the critical Josephson current
in one sample can be very different from that in another sample.
When a ferromagnet is half-metallic, the Josephson current vanishes in the absence
of spin-flip scattering at junction interfaces.
Spin-flip scattering at interfaces allows equal-spin-triplet odd-frequency Cooper pairs to penetrate into a ferromagnet.
The ratio of odd-frequency pairs to even-frequency ones
depends on the exchange potential in a ferromagnet and the spin-flip potential at interfaces.
The Josephson current recovers the self-averaging property when the fraction of
equal-spin-triplet pairs becomes large.
In half-metallic SFS junctions, all Cooper pairs have odd-frequency
symmetry. The penetration of odd-frequency pairs enhances low energy quasiparticle density
of states in a ferromagnet.
Such low energy spectra could be probed by scanning
tunneling spectroscopy and determining a nonmonotonic temperature dependence of the critical Josephson
current.
We also discuss a way to realize a $\pi$-junction by controlling magnetic moments
in ferromagnetic layers.

\begin{acknowledgments}
We acknowledge helpful discussions with J.~Aarts, T.~M.~Klapwijk, G.~E.~W.~Bauer, Yu.~V.~Nazarov,
S.~Maekawa, S.~Takahashi, A.~I.~Buzdin, A.~F.~Volkov and A.~Brinkman.
This work was partially supported by the
Dutch FOM, the NanoNed program under grant TCS7029 and Grant-in-Aid for Scientific
Research from The Ministry of Education,
Culture, Sports, Science and Technology of Japan
(Grant No. 19540352, 18043001, 17071007 and 17340106).
\end{acknowledgments}

\appendix
\section{fluctuations of Josephson current}
\begin{figure}[tbh]
\begin{center}
\includegraphics[width=8cm]{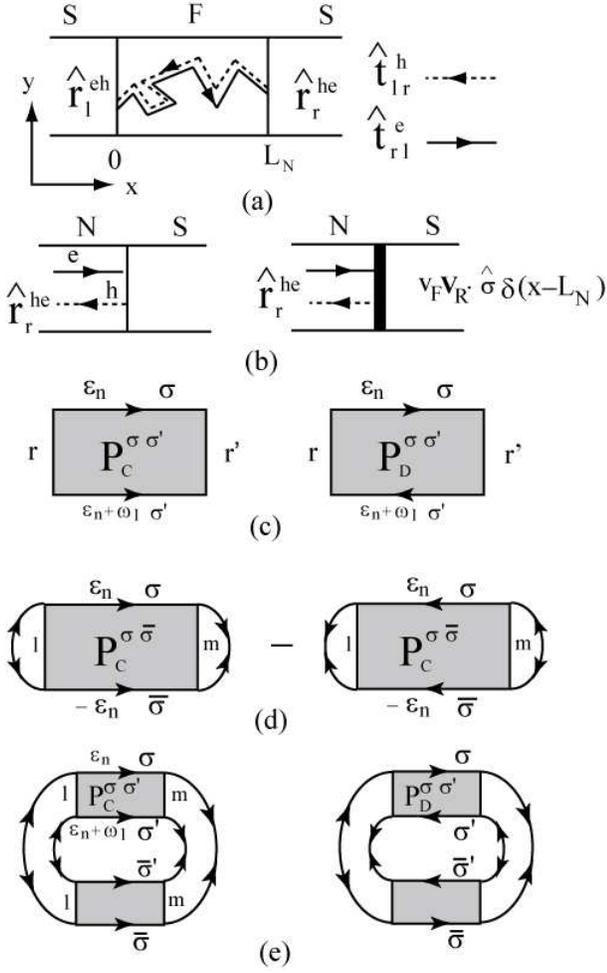}
\end{center}
\caption{ (a): a propagation process of a quasiparticle in a SFS junction.
We describe the Andreev reflection coefficients in Eq.~(\ref{formula}) by
those at an ideal normal-metal/superconductor interface as shown in (b).
(c): Cooperon and diffuson propagator.
The diagrams for Josephson current (d) and its fluctuations (e).
 }
\label{fig14}
\end{figure}

The purpose of this appendix is to explain Eq.~(\ref{dj}).
Since fluctuations of Josephson current have been calculated by the diagrammatic
expansion~\cite{altshuler,zyuzin,koyama}, we also calculate
the Josephson current in SFS junctions in the same method.
We assume that relations
$E_{Th}\ll \Delta_0, V_{ex} \ll \mu$ and $L_N\gg \ell$ are satisfied.
In the lowest coupling,
the Josephson current is given by a formula~\cite{ya01-3}
\begin{equation}
J={ie}\sum_{{l},{r}}
T\sum_{\omega_n}\textrm{Tr}
\left[
\hat{r}^{eh}_l \cdot \hat{t}^h_{lr}  \cdot \hat{r}^{he}_r  \cdot \hat{t}^e_{rl}
 - \hat{r}^{he}_l \cdot  \hat{t}^e_{lr}  \cdot \hat{r}^{eh}_r  \cdot \hat{t}^h_{rl} \right],
\label{formula}
\end{equation}
where $l$ ($r$) denotes a propagating channel at the left (right) junction interface.
In Fig.~\ref{fig14}(a), a propagation process of the first term in Eq.~(\ref{formula})
is schematically illustrated.
We calculate the transmission coefficients in a ferromagnet
such as $\hat{t}^{e}_{rl}$ and $\hat{t}^{h}_{lr}$
and Andreev
reflection coefficients  at interfaces such as $\hat{r}^{he}_r$ and $\hat{r}^{eh}_l$ by parts.
The Andreev reflection coefficients are calculated at
an ideal NS interface as shown in the left figure of Fig.~\ref{fig14}(b).
The results are given by
\begin{align}
\hat{r}^{he}_{l(r)} =& - \hat{\sigma}_2 \frac{\Delta_0}{\omega_n+\Omega_n}\;  e^{-i\varphi_{L(R)}},\\
\hat{r}^{eh}_{l(r)} =& \hat{\sigma}_2\; \frac{\Delta_0}{\omega_n+\Omega_n}  e^{i\varphi_{L(R)}},
\end{align}
where $\Omega_n=\sqrt{\omega_n^2+\Delta_0^2}$.
The effect of the exchange potential is considered through
transmission coefficients of an electron in a ferromagnet
\begin{align}
\hat{t}_{lr}^{e} = \left( \begin{array}{cc}
t^{e}_{lr}(\uparrow) & 0 \\
0 & t^{e}_{lr}(\downarrow)
\end{array}\right).
\end{align}
The transmission coefficients of a hole are defined in the same way by $e \to h$ in the
equation above.
The transmission coefficients are represented by the Green function as
\begin{align}
t^e_{rl}(\sigma)=& i e^{ik_lx_L-ik_rx_R} v_l
\iint dy_L dy_R Y^\ast_r(y_R)Y_l(y_L) \nonumber\\&\times{G}_{\omega_n}^\sigma
(x_R, y_R; x_L, y_L) \\
t^h_{lr}(\sigma)=& -i e^{-ik_lx_L+ik_rx_R} v_r
\iint dy_L dy_R Y^\ast_l(y_L)Y_r(y_R) \nonumber\\&\times{G}_{-\omega_n}^{{\sigma}}
(x_R, y_R; x_L, y_L),
\end{align}
where $Y_l(y)$ is a wave function in the $y$ direction and
$v_l=k_l/m$ with $k_l$ being
a wave number in the $x$ direction on the Fermi surface in the $l$ th propagating channel.
In above expression, we have assumed that
two ideal lead wires are attached to the both sides of a diffusive
ferromagnet, and $x_L<0$ and $(x_R>L_N)$ are taken to be in the lead wires.
The Green function is given by
\begin{align}
{G}_{\omega_n}^\sigma(\boldsymbol{r},\boldsymbol{r}')
=\frac{1}{(2\pi)^2}\!\!\int \!\!\! \frac{ d\boldsymbol{k}\; e^{i\boldsymbol{k}\cdot(\boldsymbol{r}-\boldsymbol{r}')}}
{i\omega_n\! - \xi_k\! + V_{ex} s\! + \frac{i}{2\tau}\textrm{sgn}(\omega_n)},
\end{align}
where $\tau$ is the elastic mean free time, $\xi_k= \boldsymbol{k}^2/m -\mu$, and
$s=1(-1)$ for $\sigma=\uparrow(\downarrow)$.
An ensemble average of transmission coefficients is
calculated by the diagrammatic expansion
\begin{align}
&\sum_{lr}\left\langle t_{lr}^e(\sigma) t_{rl}^h({\sigma}') \right\rangle \nonumber\\
&= \frac{v_F^2}{2} \int_0^W \!\!\!\!\!\! dy_L \!\! \int_0^W\!\!\!\!\!\!dy_R 
P_C^{\sigma {\sigma}'}(L+\delta,y_R;-\delta,y_L;2\omega_n), \label{a8}\\
&P_C^{\sigma {\sigma}'}(\boldsymbol{r}, \boldsymbol{r}')=
\langle G_{\omega_n}^\sigma(\boldsymbol{r}, \boldsymbol{r}')G_{-\omega_n}^{\sigma'}
(\boldsymbol{r}, \boldsymbol{r}')\rangle,
\end{align}
where $P_C^{\sigma {\sigma}'}$ is the Cooperon propagator which satisfies the equation
\begin{align}
&\left[ |\omega_l| - 2i V_{ex} s (1-\delta_{\sigma,\sigma'}) -D \nabla^2 \right]
 P_{C,D}^{\sigma {\sigma}'}(\boldsymbol{r},\boldsymbol{r}';\omega_l)\nonumber\\&= 2\pi N_0
\delta(\boldsymbol{r}-\boldsymbol{r}').
\end{align}
Since $\mu \gg V_{ex}$, the diffusion constant $D$, the Fermi velocity $v_F$
and the density of states at the Fermi energy $N_0$ do not depend on spin directions.
In Fig.~\ref{fig14}(c), we illustrate the Cooperon and diffuson propagator, where
$\omega_l=2\pi l T$ is a boson Matsubara frequency.
In Fig.~\ref{fig14}(d) we show two diagrams which contribute to the Josephson current.
The left (right) diagram in Fig.~\ref{fig14}(d) corresponds to the first (second) term of
Eq.~(\ref{formula}).
Only $P_C^{\uparrow\downarrow}$ and $P_C^{\downarrow\uparrow}$ contribute
to the Josephson current because the Andreev reflection coefficients
are off-diagonal in spin space.
To calculate the Cooperon propagator, we solve the diffusion equation
with appropriate boundary conditions~\cite{ya01-2}
\begin{align}
&D\nabla^2h_\lambda(\boldsymbol{r})=\lambda h_\lambda(\boldsymbol{r}),\\
&\left.h_\lambda(\boldsymbol{r})\right|_{x=0,L_N}=0,\\
&\left.\frac{\partial h_\lambda(\boldsymbol{r})}{\partial y}\right|_{y=0,W}=0.
\end{align}
The Cooperon propagator is represented by using
wave functions and their eigen values of the diffusion equation.
The results are
\begin{align}
P_{C}^{\sigma {\sigma}'}&(\boldsymbol{r},\boldsymbol{r}';2\omega_n)
=2 \pi N_0 \sum_{n=1}^{\infty}\sum_{m=0}^{\infty}\left(\frac{2}{L_N}\right)B_m \nonumber\\
\times &\frac{
\sin(p_nx)\sin(p_nx') \cos(\nu_m y)\cos(\nu_m y')}
{2 \left\{|\omega_n| - i V_{ex} s (1-\delta_{\sigma,\sigma'})\right\} + D( p_n^2 + \nu_m^2)}, \\
p_n =& \frac{n\pi}{L_N},\; \nu_m = \frac{m\pi}{W},\\
B_m=&\left\{ 
\begin{array}{cc} 
1/W & \text{for}\; m=0\\
2/W & \text{for}\; m \neq 0,
\end{array}
\right..
\end{align}
By substituting the above results into Eq.~(\ref{a8}), we arrive at
\begin{align}
\sum_{lr}&\left\langle t_{lr}^e(\sigma) t_{rl}^h(\bar{\sigma}) \right\rangle
=g_N\left( \frac{\eta^{\sigma\bar{\sigma}}}{\sinh\eta^{\sigma\bar{\sigma}}} \right),
\label{avet}\\
g_N =&\pi N_0 D W/L_N,\\
\eta^{\sigma {\sigma}'}=&\left\{
\begin{array}{cl}
\sqrt{ \frac{2|\omega_n|-2isV_{ex}}{E_{Th}}} & \sigma'=\bar{\sigma}\\
\sqrt{ \frac{2|\omega_n|}{E_{Th}} } & \sigma'={\sigma}
\end{array}\right. ,\label{etadef}
\end{align}
where $(2e^2/h)g_N$ is the conductance of a ferromagnet,
$\bar{\sigma}$ denotes the opposite spin state of $\sigma$,
 and
integration in the $y$ direction in Eq.~(\ref{a8}) is carried
out at $\delta=\ell/\sqrt{2}$.
The Josephson current becomes
\begin{align}
\langle J\rangle =& 2e\,g_N \sin\varphi \, T\sum_{\omega_n} \sum_\sigma 
 \frac{\eta^{\sigma\bar{\sigma}}}{\sinh\eta^{\sigma\bar{\sigma}}} \!\!
\left[ \frac{\Delta_0}{\omega_n+\Omega_n} \right]^2.
\end{align}
By substituting equations
\begin{align}
&\sum_{\sigma}  \frac{\eta^{\sigma\bar{\sigma}}}{\sinh\eta^{\sigma\bar{\sigma}}} 
=4 \sqrt{\frac{2V_{ex}}{E_{Th}}} e^{-\frac{L_N}{\xi_h}}\sin\left(\frac{L_N}{\xi_h} +\frac{\pi}{4}\right),\\
&T\sum_{\omega_n} \left( \frac{\Delta_0}{\omega_n+\Omega_n} \right)^2=\frac{2\Delta_0}{3\pi},
\end{align}
into above expression, the 
Josephson current at $T=0$ results in
\begin{align}
\langle J\rangle =& \frac{16\sqrt{2}}{3\pi} e \;g_N\;  \;\Delta_0 
\sqrt{\frac{V_{ex}}{E_{Th}}} e^{-\frac{L_N}{\xi_h}}
\sin\left(\frac{L_N}{\xi_h} +\frac{\pi}{4}\right) \nonumber\\
&\times \sin\varphi. 
\end{align}
This expression is also valid for $V_{ex} \lesssim \Delta_0$ because the relation 
$V_{ex} \gg \Delta_0$ was not explicitly used in the derivation.

In Fig~\ref{fig14}(e), we show two typical diagrams for fluctuations.
Not only $P^{\sigma \bar{\sigma}}$ but also
$P^{\sigma {\sigma}}$ contributes to fluctuations.
The Cooperon $P^{\sigma \bar{\sigma}}$ behaves like $e^{-L_N/\xi_h}$ similar
to the Josephson current. On the other hand, $P^{\sigma {\sigma}} \sim e^{-L_N/\xi_T}$.
Thus amplitude of fluctuations in SFS junctions is almost
the same as that in SNS junctions.
Our approach, however, is not suitable for calculating fluctuations because
a number of diagrams contributes to fluctuations in addition to Fig.~\ref{fig14}(e).
Here we present the result for SNS junctions at $T=0$ and $\varphi=\pi/2$ obtained
by a slightly different approach~\cite{koyama}
\begin{align}
\delta J = \sqrt{ \frac{\pi}{6}} eE_{Th} \sqrt{\frac{W}{L_N}}.
\end{align}
The fluctuations in SFS junctions are given by $\delta J/\sqrt{2}$ because
 contribution of $P^{\sigma, \bar{\sigma}}$ is negligible for $V_{ex}\gg E_{Th}$.
Thus the ratio is described by
\begin{align}
\frac{\delta J}{\langle J \rangle}\sim
\frac{\sqrt{6\pi^3}}{16}
\sqrt{\frac{L_N}{W}}\frac{1}{k_F\ell}
{e^{L_N/\xi_h}} \frac{ E_{Th}^{3/2}}{\Delta_0 \sqrt{V_{ex}}}.
\end{align}
Since $P^{\sigma {\sigma}} \sim e^{-L_N/\xi_T}$, temperature dependence of
fluctuations is also expected to be $ e^{-L_N/\xi_T}$.
Thus we arrive at Eq.~(\ref{dj}).
In a recent paper, mesoscopic fluctuations of the Josephson current were calculated 
within the quasiclassical Green function technique~\cite{houzet2}. In this approach 
the fluctuations are slightly larger than those within the 
diagrammatic expansion~\cite{altshuler,koyama}.
The difference may stem from the proximity effect
on electronic structure in a normal metal such as the minigap in the quasiparticle
density of states.
In the diagrammatic expansion, such effect is not taken into account.

\section{negative Josephson coupling}
Here we express the Josephson current in S/HM/S junctions with spin-active
interface on the basis of the diagrammatic expansion.
In a half metal, we assume that the magnetic moment is parallel to $\boldsymbol{e}_3$.
Thus transmission coefficients of an electron become
\begin{align}
\hat{t}^{e}_{rl}=\frac{\hat{\sigma}_0+\hat{\sigma}_3}{2}t^{e}_{rl}(\uparrow)
\end{align}
 because electric structure for $\downarrow$ spin is insulating in a half metal.
Transmission coefficients of a hole are defined in the same way by $e \to h$.
Andreev reflection coefficients $\hat{r}_r^{he}$ and $\hat{r}_r^{eh}$
in Eq.~(\ref{formula}) are calculated at
a normal-metal/ superconductor interface at which a spin-flip potential
$v_F \boldsymbol{V}_{R}\cdot \hat{\boldsymbol{\sigma}} \delta(x-L_N)$ is
introduced as shown in
the right figure of Fig.~\ref{fig14}(b). Andreev reflection coefficients
$\hat{r}_l^{he}$ and $\hat{r}_l^{eh}$ are also calculated
at a normal-metal/ superconductor interface at which a spin-flip potential
$v_F \boldsymbol{V}_{L}\cdot \hat{\boldsymbol{\sigma}} \delta(x)$ is considered.
 We assume that
$\boldsymbol{V}_{L(R)}=\sum_{k=1}^{3} V_{L(R)}^{(k)}\boldsymbol{e}_k$ and
 $|\boldsymbol{V}_{L}|=|\boldsymbol{V}_{R}|=V_S$.
The calculated results of Andreev reflection coefficients are given by
\begin{align}
\hat{r}^{he}_{l}=& -Q_{l} \hat{\sigma}_2 \left[ A_{l} \hat{\sigma}_0 + i B_{l}
 \boldsymbol{V}_{L} 
\cdot \hat{\boldsymbol{\sigma}} \right]e^{-i\varphi_{L}}, \\
\hat{r}^{eh}_{l}=& Q_{l}  
\left[ A_{l} \hat{\sigma}_0 + i B_{l}\boldsymbol{V}_{L} 
\cdot \hat{\boldsymbol{\sigma}} \right]\hat{\sigma}_2e^{i\varphi_{L}}, \\
Q_{l}=& \frac{\Delta_0}{ A_{l}^2+ V_S^2 B_{l}^2}\frac{q_{l}^2}{2},\\
A_{l}=&  -\Omega_nV_S^2+ q_{l}^2\frac{(\Omega_n+\omega_n)}{2},\\
B_{l}=& q_{l}(\Omega_n+\omega_n),
\end{align}
where $q_l=k_l/k_F>0$ are normalized wave number of the $l$ th propagating channel 
in the current direction.
Andreev reflection coefficients at the right interface 
$\hat{r}^{he}_{r}$ and $\hat{r}^{eh}_{r}$ are 
also obtained by $l \to r$, $\boldsymbol{V}_{L} \to \boldsymbol{V}_{R}$, and
  $\varphi_{L} \to \varphi_{R}$ in above expression.
Since the half metal is in the diffusive transport regime,
transmission coefficients across the half metal, namely
$\hat{t}^e_{lr}$, $\hat{t}^e_{rl}$, $\hat{t}^h_{lr}$, and $\hat{t}^h_{rl}$
are independent of propagating channels $l$ and $r$.
Thus average of the Andreev reflection coefficients over all propagating
channels contribute to the Josephson current~\cite{ya01-2}.
We define such Andreev reflection coefficients as
\begin{align}
\hat{r}^{he}_{l(r)}=& \frac{1}{N_c} \sum_{l(r)} \hat{r}^{he}_{l(r)},\\
=& - \hat{\sigma}_2 \left[ a \hat{\sigma}_0 + i b \boldsymbol{V}_{L(R)}
\cdot \hat{\boldsymbol{\sigma}}\right]e^{-i\varphi_{L(R)}}, \\
\hat{r}^{eh}_{l(r)}=& \frac{1}{N_c} \sum_{l(r)} \hat{r}^{he}_{l(r)},\\
=&  \left[ a \hat{\sigma}_0 + i b \boldsymbol{V}_{L(R)}
\cdot \hat{\boldsymbol{\sigma}}\right]\hat{\sigma}_2e^{i\varphi_{L(R)}},
\end{align}
where $N_c=Wk_F/\pi$ is the number of propagating channels at Fermi energy,
$a$ and $b$ are real numbers depending only on $\omega_n$, $\Delta_0$, and $V_S$.
A part of Eq.~(\ref{formula}) becomes
\begin{align}
I_1=&\sum_{{l},{r}} \left\langle
\textrm{Tr}
[
\hat{r}^{eh}_l \cdot \hat{t}^h_{lr}  \cdot \hat{r}^{he}_r  \cdot \hat{t}^e_{rl}]
\right\rangle, \\
=& - \sum_{{l},{r}} \left\langle t_{lr}^h(\uparrow) t_{rl}^e(\uparrow) \right\rangle
\frac{e^{i\varphi}}{4}
\textrm{Tr} \left[(a+ib \boldsymbol{V}_R\cdot \hat{\boldsymbol{\sigma}})\right.\nonumber\\
&\times\left.
(\hat{\sigma}_0-\hat{\sigma}_3)
(a+ib \boldsymbol{V}_L\cdot \hat{\boldsymbol{\sigma}})
(\hat{\sigma}_0+\hat{\sigma}_3)\right],\\
=&e^{i\varphi}b^2g_N \frac{\eta^{\uparrow\uparrow}}{\sinh\eta^{\uparrow\uparrow}} \nonumber\\
\times&\left[\boldsymbol{V}_L\cdot\boldsymbol{V}_R - {V}_L^{(3)}
{V}_R^{(3)}  + i \boldsymbol{e}_3 \cdot
(\boldsymbol{V}_R\times\boldsymbol{V}_L)
\right],
\end{align}
where we used Eq.~(\ref{avet}).
In the same way, we obtain
\begin{align}
I_2=&\sum_{{l},{r}} \left\langle
\textrm{Tr}
[
\hat{r}^{eh}_r \cdot \hat{t}^h_{rl}  \cdot \hat{r}^{he}_l  \cdot \hat{t}^e_{lr}]
\right\rangle, \\
=&e^{-i\varphi}b^2g_N \frac{\eta^{\uparrow\uparrow}}{\sinh\eta^{\uparrow\uparrow}} \nonumber\\
\times&\left[\boldsymbol{V}_L\cdot\boldsymbol{V}_R - {V}_L^{(3)}
{V}_R^{(3)}  - i \boldsymbol{e}_3 \cdot
(\boldsymbol{V}_R\times\boldsymbol{V}_L)
\right].
\end{align}
As a result, the expression for the Josephson takes the form
\begin{align}
\langle J \rangle \approx& - J_1
\left[ \left(\boldsymbol{V}_L\cdot\boldsymbol{V}_R - {V}_L^{(3)}
{V}_R^{(3)}
\right)\sin \varphi \right.\nonumber\\
& \left. + \boldsymbol{e}_3 \cdot
(\boldsymbol{V}_L\times\boldsymbol{V}_R) \cos\varphi\right],\label{jhm2}\\
J_1=&2{e} g_N T\sum_{\omega_n}\frac{\eta^{\uparrow\uparrow}}{\sinh \eta^{\uparrow\uparrow}}
b^2 > 0.
\end{align}
The Josephson current is zero in the absence of spin-flip scattering
at the interface (i.e., $\boldsymbol{V}_L=\boldsymbol{V}_R=0$).
We note that the ratio ${\eta^{\uparrow\uparrow}}/{\sinh \eta^{\uparrow\uparrow}}$ rapidly decreases
to zero for $\omega_n/E_{Th} \gg 1$, whereas $\omega_n$ dependence of $b$
is scaled by $\Delta_0$. For $E_{Th} \ll \Delta_0$, we find at $T=0$
\begin{align}
J_1=& \frac{7 \zeta(3)}{\pi} eE_{Th} \; g_N b^2,\\
b=&\frac{1}{2}\int_0^{\pi/2}\!\!\!\!d\gamma\;
\frac{\cos^5\gamma}{\left(V_S^2+\frac{1}{4}\right)\cos^4\gamma - V_S^2 \cos^2\gamma
+V_S^4}.
\end{align}
Although Eq.~(\ref{jhm2}) describes well the dependence of the Josephson current on
$\boldsymbol{V}_L$ and $\boldsymbol{V}_R$, it does not
explain the nonmonotonic temperature dependence of the critical current shown in
Fig.~\ref{fig11}(b).
This is because the proximity effect on the density of states in a half
metal is not taken into account in the above estimate.


\begin{thebibliography}{99}
\bibitem{length} In two-dimensional lattices, wave function of a quasiparticle
is basically localized due to random impurity potential.
In the present parameter choice, the localization length $\xi_{AL}$
and the mean free path $\ell$ at $V_{ex}=0$ are estimated about 90 and 6 lattice
constant, respectively. We have confirmed that these values are not so much
sensitive to $V_{ex}$.
The present parameter choice enables us to study transport in the diffusive transport regime
because a relation $\ell \ll L_N < \xi_{AL}$ is satisfied. Effects of the localization
on Josephson current are important when we choose $L_N \gg \xi_{AL}$~\cite{ya02-4}.
\end{thebibliography}
\end{document}